\documentclass[12pt]{article}
\usepackage{amssymb}
\usepackage{amsfonts}
\usepackage{amsmath}
\usepackage[mathscr]{eucal}
\usepackage{amssymb}
\usepackage{amsthm}
\usepackage{mathrsfs}
\usepackage{bbold}
\usepackage{bm}
\usepackage{graphicx}
\usepackage{caption}
\usepackage[english]{babel}
\usepackage[T2A]{fontenc}
\usepackage[utf8]{inputenc}
\usepackage{cite}

\newcommand{\be}{\begin{equation}}
\newcommand{\ee}{\end{equation}}
\newcommand{\bea}{\begin{eqnarray}}
\newcommand{\eea}{\end{eqnarray}}

\newcommand{\lb}{\label}
\newcommand{\p}[1]{(\ref{#1})}

\usepackage{MnSymbol}

\newcounter{rown}

\begin{document}
\begin{titlepage}
\vspace*{0.1cm}

\begin{center}
{\LARGE\bf Infinite (continuous) spin particle}

\vspace{0.3cm}

{\LARGE\bf in constant curvature space}



\vspace{1cm}

{\large\bf I.L.\,Buchbinder$^{1,2,3}$\!\!,\,\,
S.A.\,Fedoruk$^1$\!,\,\,  A.P.\,Isaev$^{1,5}$\!,\,\,  V.A.\,Krykhtin$^{2,4}$}

\vspace{1cm}

\ $^1${\it Bogoliubov Laboratory of Theoretical Physics,
Joint Institute for Nuclear Research, \\
141980 Dubna, Moscow Region, Russia}, \\
{\tt buchbinder@theor.jinr.ru, fedoruk@theor.jinr.ru,
isaevap@theor.jinr.ru}

\vskip 0.5cm

\ $^2${\it Department of Theoretical Physics,
Tomsk State Pedagogical University, \\
634041, Tomsk, Russia}, \\
{\tt joseph@tspu.edu.ru}

\vskip 0.5cm

\ $^3${\it Tomsk State University of Control Systems
and Radioelectronics,\\
\it Lenin Av.\, 40, 634050, Tomsk, Russia}\\

\vskip 0.5cm

\ $^4${\it National Research Tomsk State  University,}\\{\it Lenin
Av.\ 36, 634050, Tomsk, Russia},\\
{\tt krykhtin@tspu.edu.ru}

\vskip 0.5cm

\ $^5${\it Faculty of Physics, Lomonosov Moscow State University,
119991 Moscow, Russia}

\end{center}

\vspace{2cm}

\nopagebreak

\begin{abstract}
\noindent We present a new particle model that generalize for constant curvature space an infinite spin particle in flat space.
The model is described by commuting Weyl spinor additional coordinates.
It proved that such a model is consistent only in external gravitational field corresponding to the constant curvature spaces.
Full set of the first-class constraints in the de Sitter and anti-de Sitter spaces is obtained.
\end{abstract}

\vspace{3cm}


\phantom{Keywords: }

\newpage

\end{titlepage}
\setcounter{footnote}{0}
\setcounter{equation}{0}

\section{Introduction}
Models of relativistic particles in external fields are the useful tools for studying and clarifying the various properties
of more complicated field or string theories at classical and quantum levels. There is a giant literature  devoted to formulation of such models and
exploring the various aspects of their gauge structure, (super)symmetry, dynamics, quantization and applications for calculation of the quantum effects within
the world-line formalism. In this paper we propose a new model of a $4D$ relativistic particle in a gravitational background, therefore for the review we note only those works that are related to the description of (super) particles in various external (super) gravitational fields (see e.g. \cite{BCL,BMSS,R,LM,HPPT,BSh,RH,KYa,BLPS,GG,GG1,BCL1,BBCL,ABT,U} and the references therein).

As well known from pioneer papers by Wigner \cite{Wig} and Bargmann and Wigner \cite{BargWig}, the irreducible unitary representations of the Poincar\'{e} group are associated with elementary particles. The physically noncontradictory irreducible representations are divided into finite spin massive and massless, in its turn, the massless ones are divided into representations with finite helicities and infinite helicity representations, last are called the infinite or continuous spin representations. Massive and finite helicity representations are broadly used to describe the standard massive and massless elementary particles; as for the infinite spin representations, their physical meaning is unclear yet. Nevertheless, the infinite spin representations attract definite attention due to the hypothetical possibility to construct on such a ground a consistent massless higher spin interacting theory. At present, there is a large enough activity devoted to various aspects of describing infinite spin representations within field theory (see e.g. the review \cite{BekSk} and the
further development in the recent papers \cite{BekMur,SchTor,Met,Zin1,Zin,AG,BKT,M,BIF,BFIK} and the reference therein).

As we pointed out, many features of complicated field theories can be studied in simplified forms within appropriate particle models. The corresponding particle models associated with massive or massless helicity irreducible representations are well known and broadly used for different purposes. As to the particle models related to infinite spin representations, such models are practically unknown. It seems, the only exclusion is our work \cite{BFIR} where a dynamical model of infinite spin particle in $4D$ flat space was constructed and explored. In this paper we generalize the results of the above work to the particle in constant curvature space. However, before to go on, we should make some remarks.

First, many years ago, Bargmann and Wigner \cite{BargWig} shown that a description of infinite spin particle in flat space besides space-time coordinates must include some auxiliary "intrinsic" coordinates which can e.g. be vector or spinor. It is natural to assume that a description in a curved space-time must also use auxiliary coordinates that, in this case, must somehow be consistent with geometric structure of space-time. For example, the auxiliary spinor coordinate should be spinor under local Lorentz transformations.

Second, notion of infinite spin relates to irreducible representations just of the Poincar\'{e} group and does not admit a direct generalization to arbitrary curved space-time, even to non-zero constant curvature space-time. Therefore, when we say about infinite spin particle model in curved space-time we take into account the general covariant model with the same number of degrees of freedom like in the corresponding particle model in flat space and with known flat limit A number of degrees of freedom in the infinite spin particle model in flat space is managed by system of first class constraints. Therefore, construction of infinite spin particle models means in essence finding the general covariant first class constraints.

Third, we begin constructing an infinite spin particle model in an arbitrary curved space-time, generally speaking with torsion and looking for general covariant generalization of the flat space constraints that form a closed algebra in terms of Poisson brackets. As we will see, such consistent requirements are extremely non-trivial and yield that the space-time must be only constant curvature space with vanishing torsion\footnote{A similar problem was considered in the higher spin theory in \cite{Buchbinder:2011vp}.}.

As far as we know, model of infinite spin particle in curved space-time never been before studied.
The only what was presented in literature is discussion of some aspects of infinite spin field theories in AdS space, basically in context of their Lagrangian formulation. In principle, such field models in curved space can be constructed or upon quantization of the corresponding particle (actually this problem never been studied) or using the various methods independently of quantization of  particle models (see \cite{BekMur,SchTor,Met,Zin1,Zin,AG,BKT,M,BIF,BFIK,BFIK-1}).  
Before to go on, we would like to dwell on some question that can arise when trying to find relation between infinite spin particle model in (A)dS space, presented in this paper, and infinite spin field model in AdS space. 
As we will see, the infinite spin particle model is well defined and completely consistent in constant curvature space.  
Corresponding field model constructed in AdS space (see \cite{Met,BFIK-1}) is also well defined as a gauge field theory. 
However, when considering the field model some specific question arises. 
Since the irreducible massless representation with infinite spin in the AdS space, unlike the Minkowski space, is not defined, 
one can ask how such an infinite spin field relates to the  massless representations of the AdS symmetry group. The question has no direct relation both to particle model constructed here and to Lagrangian field theory considered in \cite{Met,BFIK-1}. 
Nevertheless, we assume that discussion of this point is useful. 
This question was considered in the review \cite{BekSk}, where some thoughts were made, although a complete answer was not given. To be precise, such a  question was noted in \cite{BekSk} as one of the open issues in this area. 
At best, we can expect that the field, called the infinite spin field in AdS space is at least decomposed 
into representations of the AdS group, including all spins. This consideration in fact was implemented in the works \cite{Met} when constructing the Lagrangian formulation. The infinite spin field Lagrangian was formulated as an infinite sum of Fronsdal Lagrangians in AdS space by adding to this sum of Lagrangians and gauge transformations the infinite number of new terms mixing the fields with different spins. It is important to emphasize that such a mixture is stipulated both by parameter ${\bm \mu}$ characterizing the infinite spin representation and by the parameter $\kappa$ characterizing the AdS space. In the other words, at ${\bm \mu} =0$, the infinite spin Lagrangian on AdS does not decay into a sum of Lagrangians corresponding to AdS irreducible representations with given spins. In this context, the infinite spin field in AdS space means a special mixture of infinite number of irreducible AdS fields with all spins. However,  this is completely consistent gauge field model that leads in the flat limit to correct model of an infinite spin field in Minkowski space. As we will see, in the infinite spin particle model in the space of constant curvature a correct flat limit is also exists.

The paper is organized as follows. Section\,2 is a brief review of our Lagrangian construction for infinite spin particle model in flat space \cite{BFIR} focusing on structure of first class constraints. In section\,3 we discuss a generic scheme to construct a Lagrange formulation in arbitrary curved space-time with torsion. The section\,4 is devoted to deriving curved space-time generalization of the flat space constraints and finding the conditions yielding closed algebra of constraints in curved space. It is proved here that a closed algebra the constraints is possible only in torsion free constant curvature space-time.
Section\,5 summarizes the results obtained.

\vspace{0.5cm}

\setcounter{equation}{0}

\section{Infinite spin particle in flat space}

Lagrangian formulation of free infinite spin particle in Minkowski space has been constructed in the work \cite{BFIR} within the first-order formalism using the spinor auxiliary coordinates. As it is known (see e.g. \cite{HPPT}), the Lagrangian for any free relativistic particle is basically defined in the first order formalism  by a system of constraints. In the case under consideration, the Lagrangian is written as follows
\begin{equation}
\label{L-sp}
L_{flat} \ = \ p_a \dot x^a \ +  \ \pi_{\alpha} \dot \xi^{\alpha}  \ +  \ \bar\pi_{\dot\alpha} \dot {\bar\xi}^{\dot\alpha}  \ +  \ e_0 f_0 \  + \
e f \  + \  \tilde e \tilde f\  + \  e_u u\,,
\end{equation}
where $e_0(\tau)$, $e(\tau)$, $\tilde e(\tau)$, $e_u(\tau)$ are the Lagrange multipliers for
the Bargmann-Wigner constraints
\begin{eqnarray}
f_0 &:=& p_a p^a \ \approx \ 0 \,, \label{const-sp}\\ [5pt]
f &:=& p_a \left(\xi\sigma^a \bar\xi \right)-\bm{\mu}  \ \approx \ 0  \,,  \label{const-sp-1}\\ [5pt]
\tilde f &:=& p_a \left(\bar\pi\tilde\sigma^a \pi \right) -\bm{\mu} \ \approx \ 0  \,,  \label{const-sp-2}\\ [5pt]
u &:=& N -\bar N \ \approx \ 0 \,,  \label{const-sp-3}
\end{eqnarray}
where
\begin{equation}
\label{N}
N:=\xi^{\alpha}\pi_{\alpha}\,,\qquad \bar N:=\bar\pi_{\dot\alpha} {\bar\xi}^{\dot\alpha}\,.
\end{equation}
The parameter $\bm{\mu}\in\mathbb{R}$ in \p{const-sp-1}, \p{const-sp-2} is a nonzero
constant, $x^a$, $p_a$ are coordinates and momenta of the relativistic particles,
$\xi^\alpha$,\, $\bar\xi^{\dot\alpha}$ are the spinor auxiliary coordinates,
$\pi_\alpha$,\, $\bar\pi_{\dot\alpha}$ are their momenta and $\tau$ is an evolution parameter: $x^a=x^a(\tau)$,
$p_a=p_a(\tau)$, etc. We use the standard notation for
$\tau$-derivatives: $\dot x^a$, $\dot \xi^{\alpha}$,
$\dot {\bar\xi}^{\dot\alpha}$. Constraints \p{const-sp-1}, \p{const-sp-2} are imposed to get the infinite-spin irreducible representation of the Poincar\'{e} group. We stress that in the case when one uses the spinor auxiliary coordinates, we should introduce the constraint $u\approx 0$ \p{const-sp-3}
that commutes with all Casimir operators of the Poincar\'{e} group. This constraint provides an equal number of dotted and undotted spinor indices
in the dynamical invariants and hence guarantees a possibility to use in such quantities only vector indices (see details in \cite{BFIR}).

The non-vanishing Poisson brackets  of the canonical momenta $p_a$,
$\pi_\alpha$, $\bar\pi_{\dot\alpha}$ and the four-vector coordinates $x^a$ and
auxiliary the Weyl spinors $\xi^\alpha$, $\bar\xi^{\dot\alpha}$ are
\begin{equation}
\label{PB0}
\left\{ x^a, p_b \right\}=\delta^a_b\,,\qquad
\left\{\xi^\alpha, \pi_\beta \right\}=\delta^\alpha_\beta\,,\qquad
\left\{\bar\xi^{\dot\alpha}, \bar\pi_{\dot\beta} \right\}=\delta^{\dot\alpha}_{\dot\beta}\,.
\end{equation}
Therefore, the non-vanishing Poisson brackets of the constraints \eqref{const-sp}-\eqref{const-sp-3} have the form
\begin{equation}
\label{algebra}
\left\{ f, \tilde f \right\} \ = \ -K f_0\,,
\end{equation}
where
\be\lb{K}
K \ := \ N +\bar N
\ee
and $N$ and $\bar N$ are given by \p{N}.
Thus, the constraints \eqref{const-sp}-\eqref{const-sp-3} form closed Poisson brackets algebra.

Main purpose of the paper is to generalize this Lagrangian approach to the case of the particle in a curved space-time with preservation of the  same number of degrees of freedom like in the flat space. The basic requirements for such a construction are:

$\bullet$ General covariant generalization of the constraints \eqref{const-sp}-\eqref{const-sp-3} with given flat limit.

$\bullet$ Closure of the algebra of new constraints. One can expect that this condition imposes restriction on the space-tim geometry.

In the next section we will consider a realization of the above requirements.

\vspace{0.5cm}

\setcounter{equation}{0}

\section{Generalization to curved space-time}

Now $x^\mu(\tau)$ are local coordinates in a curved space, where $\mu$ are vector indices. The two-component quantities $\xi^\alpha$, $\bar\xi^{\dot\alpha}$, $\pi_\alpha$, $\bar\pi^{\dot\alpha}$ are the spinors in tangent space with the spinor indices which are raised and lowered by $\epsilon_{\alpha\beta}$ and $\epsilon^{\alpha\beta}$. It is assumed from the very beginning that the space-time geometry
is described by metric and general spin connection. We also assume that the Lagrangian for particle in an external gravitational background is constructed according to generical scheme in the form $L=p\,\dot{x}+\pi \dot{\xi}+ \bar\pi\dot{\bar\xi}+(\mbox{terms containing the constraints})$. The canonically conjugate coordinate $x^{\mu}$, $\xi^{\alpha}$, ${\bar\xi}^{\dot\alpha}$ and momenta  $p_{\mu}$, $\pi_{\alpha}$, ${\bar\pi}_{\dot\alpha}$ satisfy the standard Poisson brackets (sf. \p{PB0})
\begin{equation}
\label{PB-sp}
\left\{ x^\mu, p_\nu \right\}=\delta^\mu_\nu\,,\qquad
\left\{\xi^\alpha, \pi_\beta \right\}=\delta^\alpha_\beta\,,\qquad
\left\{\bar\xi^{\dot\alpha}, \bar\pi_{\dot\beta} \right\}=\delta^{\dot\alpha}_{\dot\beta}\,.
\end{equation}
Local coordinates of the momentum $p_\mu$ is a world vector.

The Lagrangian $L$ and the constraints by definition should be the scalars under the general coordinate transformations. However, since the spinors are defined under the local Lorentz transformations in the tangent space, their $\tau$-derivatives do not satisfy the true spinor transformation laws, it is convenient to use the covariant $\tau$-derivatives: $D\xi^\alpha =\dot \xi^\alpha+\dot x^\mu \omega_\mu{}^{\alpha\beta} \xi_\beta$, \, $D{\bar\xi}^{\dot{\alpha}}=\dot {\bar\xi}^{\dot\alpha}+\dot x^\mu \omega_\mu{}^{\dot\alpha\dot\beta} {\bar\xi}_{\dot\beta}$, where  $\omega_\mu{}^{\alpha\beta}=\omega_\mu{}^{\beta\alpha}$ and
$\omega_\mu{}^{\dot\alpha\dot\beta}=\omega_\mu{}^{\dot\beta\dot\alpha}$ are the spinor connections. Taking into account these covariant derivatives we rewrite identically
kinetic term of the Lagrangian $p\dot{x}+\pi \dot{\xi}+\bar\pi\dot{\bar\xi}$ in the form $\mathcal{P}_\mu \dot x^\mu +\pi_{\alpha}D\xi^{\alpha}+{\bar\pi}_{\dot\alpha}D\xi^{\dot\alpha}$, where the quantity $\mathcal{P}_\mu$ is defined by the relation
\begin{equation}
\label{P}
\mathcal{P}_\mu = p_\mu  + \pi_{\alpha}\omega_\mu{}^{\alpha}{}_{\beta}\xi^{\beta} + {\bar\pi}_{\dot\alpha}\omega_\mu{}^{\dot\alpha}{}_{\dot\beta}{\bar\xi}^{\dot\beta} \,.
\end{equation}
One can prove that the quantity $\mathcal{P}_\mu$ (\ref{P}) is a world vector which is consistent with that $L$ and $\pi D\xi, \, \bar\pi D\bar\xi$ are the world scalars\footnote{This statement is not so evident from the very beginning. To show that we note that the $\xi, \pi, \bar{\xi}, \bar{\pi}$ are inert under the general coordinate transformations $\delta{x}^{\mu} = \epsilon^{\mu}(x).$ In this case, $\delta\omega_\mu{}^{\alpha}{}_{\beta}=-\partial_\mu \epsilon^\nu\omega_\nu{}^{\alpha}{}_{\beta}$ and $\delta\omega_\mu{}^{\dot{\alpha}}{}_{\dot{\beta}}=-\partial_\mu \epsilon^\nu\omega_\nu{}^{\dot{\alpha}}{}_{\dot{\beta}}$. Therefore, under the general coordinate transformations, $\delta(\omega_\mu{}^{\alpha}{}_{\beta}\dot{x}^{\mu})=\delta(\omega_\mu{}^{\dot{\alpha}}{}_{\dot{\beta}}\dot{x}^{\mu})$ = 0.
Hence, $\mathcal{P}_{\mu}$ and $p_{\mu}$ are world vectors.}.
Taking into account the above discussion, a natural generalization of the flat space infinite-spin particle Lagrangian \p{L-sp}
is written in the form
\begin{equation}
\label{L-sp-nf}
L \ = \ \mathcal{P}_\mu \dot x^\mu \ +  \
\pi_\alpha \left(\dot \xi^\alpha+\dot x^\mu \omega_\mu{}^{\alpha\beta} \xi_\beta\right) \ +  \
\bar\pi_{\dot\alpha} \left(\dot {\bar\xi}^{\dot\alpha}+\dot x^\mu \omega_\mu{}^{\dot\alpha\dot\beta} \xi_{\dot\beta}\right)  \ +  \
\lambda_0 \mathcal{F}_0 \  + \
\lambda \mathcal{F} \  + \ \tilde\lambda \tilde{\mathcal{F}} \  + \  \lambda_u \mathcal{U}\,,
\end{equation}
where the quantities $\lambda_0(\tau)$, $\lambda(\tau)$, $\tilde\lambda(\tau)$, $\lambda_u(\tau)$ are the Lagrange multipliers for
the constraints $\mathcal{F}_0\,{\approx}\,0$, $\mathcal{F}\,{\approx}\,0$, $\tilde{\mathcal{F}}\,{\approx}\,0$, $\mathcal{U}\,{\approx}\,0$.
All the constraints must be the scalars under the general coordinate transformations. We will assume that they  are functions
of the variables $x^\mu$, $\mathcal{P}_\mu$, $\xi^\alpha$, $\bar\xi^{\dot\alpha}$, $\pi_\alpha$, $\bar\pi_{\dot\alpha}$, presented in the Lagrangian \eqref{L-sp-nf}. First, the constraint $\mathcal{U}\approx0$ has the same meaning as the constraint $u\approx0$ (\ref{const-sp-3}) in flat space. The main problem is to construct all the other constraints.

For further, it is convenient to move on to the formulation of geometrical objects in local Lorentz frame. First of all one rewrites the quantity $\mathcal{P}_\mu$ (\ref{P}) in the form
\begin{equation}
\label{cal-P}
\mathcal{P}_\mu \ = \ p_\mu  \ +  \ \frac{1}{2}\, \omega_\mu{}^{ab}M_{ab} \,,
\end{equation}
where
\begin{equation}
\label{Not-sp}
M_{ab} = \xi\sigma_{ab}\pi - \bar\pi\tilde\sigma_{ab}\bar\xi\,.
\end{equation}
and
\begin{equation}
\label{omega}
\omega_\mu{}^{ab} = (\sigma^{ab})_{\alpha\beta}\omega_{\mu}{}^{\alpha\beta} - (\tilde{\sigma}^{ab})_{\dot\alpha\dot\beta}\omega_{\mu}{}^{\dot\alpha\dot\beta}
\end{equation}
The Lorentz matrix generators $\sigma^{ab}$ and $\tilde{\sigma}^{ab}$ are defined according to \cite{BuchKuz}.

Using the definition of $M_{ab}$ (\ref{Not-sp}), the commutation relations (\ref{PB-sp}) and properties of the matrices $\sigma^{ab}$ and $\tilde{\sigma}^{ab}$ one can prove the following relation in terms of Poisson brackets:
\begin{equation}
\label{M-alg}
\left\{ M_{ab} ,M_{cd} \right\} \ = \ \eta_{ac}M_{bd}+\eta_{bd}M_{ac}-\eta_{ad}M_{bc}-\eta_{bc}M_{ad}\,.
\end{equation}
As a result we obtained the standard commutation relations for Lorentz group generators that allows us to identify the $M_{ab}$ with such generators and therefore to identify the quantity $\omega_{\mu}{}^{ab}$ with Lorentz connection.

Calculation of the Poisson bracket for the quantities $\mathcal{P}_\mu$ \eqref{cal-P} yields
\begin{equation}
\label{cal-P-alg}
\left\{ \mathcal{P}_\mu ,\mathcal{P}_\nu \right\} \ = \ -\frac{1}{2}\,R_{\mu\nu}{}^{ab}M_{ab}\,,
\end{equation}
\begin{equation}
\label{R-expr}
R_{\mu\nu}{}^{a}{}_{b} \ = \ \partial_\mu\omega_\nu{}^{a}{}_b -\partial_\nu\omega_\mu{}^{a}{}_b
+[\omega_\mu,\omega_\nu]^{a}{}_b
\end{equation}
is the curvature tensor in terms of Lorentz connection. Let us introduce the Lorentz vector in the tangent space
\begin{equation}
\label{P-a}
\mathcal{P}_a \ :=  \ e^\mu{}_a\mathcal{P}_\mu \,,
\end{equation}
where $e^\mu{}_a(x)$ is an inverse matrix to the vierbein matrix $e_\mu{}^a(x)$\footnote{Standard relations are
$e_\mu{}^a e_\nu{}_a=g_{\mu\nu}$,
$e^\mu{}_a e^\nu{}^a=g^{\mu\nu}$, $e_\mu{}^a e^\nu{}_a=\delta_\mu{}^\nu$.}.
In this case the commutation relations \p{cal-P-alg} yield
\begin{equation}
\label{cal-Pa-alg}
\left\{ \mathcal{P}_a ,\mathcal{P}_b \right\} \ = \ -\frac{1}{2}\,R_{ab}{}^{cd}M_{cd} \ + \ 2\tilde T_{ab}{}^{c}\mathcal{P}_c\,,
\end{equation}
where
\begin{equation}
\label{R-expr-L}
R_{ab}{}^{cd} \ = \ e^\mu{}_a e^\nu{}_b R_{\mu\nu}{}^{cd}
\end{equation}
and
\footnote{Symmetrization and antisymmetrization are defined in the form:
$A_{(a}B_{b)}=\frac12\left(A_{a}B_{b}+A_{b}B_{a}\right)$, $A_{[a}B_{b]}=\frac12\left(A_{a}B_{b}-A_{b}B_{a}\right)$, etc.}
\begin{equation}
\label{Tt-expr}
\tilde T_{ab}{}^c  = - e^\nu{_{[a}}(\partial_\nu e^\mu{}_{b]})e_\mu{}^c
\end{equation}
is connection-free part of the torsion (see \eqref{T-expr} bellow).

\vspace{0.5cm}

\setcounter{equation}{0}

\section{Construction of the first class constraints at the curved background }

Let us construct the constraints $\mathcal{F}_0\approx0$, $\mathcal{F}\approx0$, $\tilde{\mathcal{F}}\approx0$, $\mathcal{U}\approx0$ which we use in the Lagrangian (\ref{L-sp-nf}) and which are curved space generalization of the flat space constraints \p{const-sp}-\p{const-sp-3}.

\subsection{Restrictions on space-time geometry}

Taking into account the form of the constraints \p{const-sp}, \p{const-sp-1}, \p{const-sp-2}, \p{const-sp-3} we assume that the constraint $\mathcal{U}\approx0$ should coincide with flat space constraint $u\approx0$ while the other curved space constraints should be constructed from the following quantities
\be\lb{ell}
\ell_0 := \mathcal{P}_a \mathcal{P}^a \,, \qquad
\ell := \mathcal{P}_a \left(\xi\sigma^a \bar\xi \right) \,,  \qquad
\tilde\ell := \mathcal{P}_a \left(\bar\pi\tilde\sigma^a \pi \right) \,
\ee
which are the natural covariant generalization of the flat space constraints \p{const-sp}, \p{const-sp-1}, \p{const-sp-2}.
The Poisson brackets of the quantities \p{ell} are written in the completely  covariant form
\bea
\label{PB-ll}
\left\{ \ell ,\tilde\ell \right\} & = &
- K \ell_0 \  - \
\frac{1}{2}\left(\xi\sigma^a \bar\xi \right)\left(\bar\pi\tilde\sigma^b \pi \right)R_{ab}{}^{cd} M_{cd} \  + \
2\left(\xi\sigma^a \bar\xi \right)\left(\bar\pi\tilde\sigma^b \pi \right) T_{ab}{}^c \mathcal{P}_c
\,, \\ [6pt]
\label{PB-ll0}
\left\{ \ell_0 ,\ell \right\} & = &
\left(\xi\sigma^a \bar\xi \right)\mathcal{P}^b R_{ab}{}^{cd} M_{cd} \  - \
4\left(\xi\sigma^a \bar\xi \right)\mathcal{P}^b T_{ab}{}^c \mathcal{P}_c
\,,
\\ [6pt]
\label{PB-tll0}
\left\{ \ell_0 ,\tilde\ell \right\} & = &
\left(\bar\pi\tilde\sigma^a \pi \right)\mathcal{P}^bR_{ab}{}^{cd} M_{cd} \  - \
4\left(\bar\pi\tilde\sigma^a \pi \right)\mathcal{P}^b T_{ab}{}^c \mathcal{P}_c
\,.
\eea
Here the quantity $K$ is given by \p{K} and $T_{ab}{}^c$ is the torsion tensor:
\begin{equation}
\label{T-expr}
T_{\mu\nu}{}^a= \partial_{[\mu} e_{\nu]}{}^a+\omega_{[\mu,}{}^{ab}e_{\nu]}{}_b\,,\qquad
T_{ab}{}^\mu = e^\nu{}_a e^\lambda{}_b e^\mu{}_c T_{\nu\lambda}{}^c =  -e^\nu{}_{[a}\partial_\nu e^\mu{}_{b]} -
 e^\nu{}_{[a}\omega^{\phantom{\nu}}_{\nu\, b]c} e^{\mu c}
\,,
\end{equation}
\begin{equation}
\label{T-expr1}
T_{ab}{}^c  = T_{ab}{}^\mu  e_\mu{}^c\,,\qquad
T_{ab,c}= \tilde T_{ab,c}- \omega_{[a,b]c}\,.
\end{equation}
and $\tilde T_{ab,c}$ is defined by \p{Tt-expr}.


The relations \p{PB-ll}, \p{PB-ll0}, \p{PB-tll0} show that the curved space  quantities \p{ell} can not be treated as the first-class constraints since their algebra is not closed. We believe that there are only two possible related ways of constructing a closed algebra: restrictions on space-time geometry and modification of the quantities \p{ell}. As a first step to introduce the true curved space constraints, we assume that our curved space-time is torsionless,
\begin{equation}
\label{T-0}
T_{\mu\nu}{}^a= 0\,.
\end{equation}
In this case, the spin connection $\omega_{\mu}{}^{ab}$ is expressed through the vierbein $e_{\mu}{}^{a}$ by standard way and we arrive to Riemann space with corresponding curvature tensor $R_{abcd}$.

Under assumption \p{T-0}, the relations \p{PB-ll}, \p{PB-ll0}, \p{PB-tll0} are simplified and take
the form
\bea
\label{PB-ll-T0}
\left\{ \ell ,\tilde\ell \right\} & = &
- K \ell_0 \  - \
\frac{1}{2}\left(\xi\sigma^a \bar\xi \right)\left(\bar\pi\tilde\sigma^b \pi \right)R_{ab}{}^{cd} M_{cd}
\,, \\ [6pt]
\label{PB-ll0-T0}
\left\{ \ell_0 ,\ell \right\} & = &
\left(\xi\sigma^a \bar\xi \right)\mathcal{P}^b R_{ab}{}^{cd} M_{cd}
\,,
\\ [6pt]
\label{PB-tll0-T0}
\left\{ \ell_0 ,\tilde\ell \right\} & = &
\left(\bar\pi\tilde\sigma^a \pi \right)\mathcal{P}^bR_{ab}{}^{cd} M_{cd}
\,,
\eea
Now we begin to discuss the possibilities to modify the flat space constraints. Since we assumed that the constraint $\mathcal{U}$=$u$, that is
\be\lb{U}
\mathcal{U}=\ N -\bar N \approx0\,
\ee
we can use the condition \p{U} in the expressions \p{PB-ll-T0}, \p{PB-ll0-T0}, \p{PB-tll0-T0}. One can prove that the following relation takes place
\bea
\label{ident}
\left.\left(\xi\sigma^{[a} \bar\xi \right)\left(\bar\pi\tilde\sigma^{b]} \pi \right)\right|_{\mathcal{U}\approx0}=
-\frac12\,K\,M^{ab},
\eea
where we have used the identity $\sigma^{[a}_{\alpha\dot\alpha}\sigma^{b]}_{\beta\dot\beta}=-(\sigma^{ab}\epsilon)_{\alpha\beta}\epsilon_{\dot\alpha\dot\beta}-
(\epsilon\tilde\sigma^{ab})_{\dot\alpha\dot\beta}\epsilon_{\alpha\beta}$. Further we will apply the identity \p{ident} to analyse the possible modifications of the relations \p{PB-ll-T0}, \p{PB-ll0-T0}, \p{PB-tll0-T0} to define the true curved space constraints forming a closed algebra.

Direct calculations show that under identity \p{ident},
the Poisson bracket \p{PB-ll-T0} has the form:
\be\lb{PB-ll-T0a}
\left.\left\{ \ell ,\tilde\ell \right\}\right|_{N \approx\bar N} =
- K \left(\ell_0 \  - \
\frac{1}{4}\,R_{ab}{}^{cd} M^{ab} M_{cd}\right)
\,.
\ee
Therefore, as first approximation in modifying the constraint \p{const-sp} is naturally to take instead of $\ell_0\approx0$ the quantity in the right side of \p{PB-ll-T0a}:
\be\lb{ell0-1mod}
\ell_0  -
\frac{1}{4}\,R_{ab}{}^{cd} M^{ab} M_{cd} \ \approx \ 0\,.
\ee
To examine, if the quantity \p{ell0-1mod} to be an appropriate constrains, let us consider the Poisson brackets of this quantity with $\ell$ and $\tilde\ell$.
The result of calculations has the form
\begin{eqnarray}
\label{PB-1F0-l}
\left\{ \ell_0 -  \frac{1}{4}\,{R}_{ab}{}^{cd} M^{ab}M_{cd} \, , \,
\ell \right\} &=&
2\left(\xi\sigma^a \bar\xi \right)\mathcal{P}^b R_{ab}{}^{cd} M_{cd}
\ - \ \frac{1}{4}\left(\xi\sigma^h \bar\xi \right)\mathcal{D}_h {R}_{ab}{}^{cd}M^{ab}M_{cd},
\\ [6pt]
\label{PB-1F0-tl}
\left\{\tilde\ell_0 -  \frac{1}{4}\,{R}_{ab}{}^{cd} M^{ab}M_{cd} \, , \,
\tilde\ell \right\} &=&
2\left(\bar\pi\tilde\sigma^a \pi \right)\mathcal{P}^bR_{ab}{}^{cd} M_{cd}
\ - \ \frac{1}{4}\left(\bar\pi\tilde\sigma^h \pi \right)\mathcal{D}_h {R}_{ab}{}^{cd}M^{ab}M_{cd},
\end{eqnarray}
where
\be
\label{D-cur}
\mathcal{D}_h {R}_{ab}{}^{cd} =
\partial_h {R}_{ab}{}^{cd}
+\omega_{h,a}{}^g{R}_{gb}{}^{cd} +\omega_{h,b}{}^g{R}_{ag}{}^{cd}
+\omega_{h,}{}^c{}_g{R}_{ab}{}^{gd}
+\omega_{h,}{}^d{}_g{R}_{ab}{}^{cg} \ee and $\partial_a =
e^\mu{}_{a}\partial_\mu$,
$\omega_{a,bc}=e^\mu{}_{a}\omega_{\mu\,bc}$. We see that the above
relations do not form a closed algebra in general case. Therefore,
we need some restrictions on the geometry.

To clarify what kind of restrictions we should impose, we turn
attention that only the quantities $\ell$ and $\tilde\ell$ are
linear in $\mathcal{P}_a$. The right-hand sides of \p{PB-1F0-l},
\p{PB-1F0-tl} contain the terms linear in $\mathcal{P}_a$ and the
terms that do not contain the $\mathcal{P}_a$. But the expressions
$\ell$ and $\tilde\ell$ involve vector contractions $\mathcal{P}_a
\left(\xi\sigma^a \bar\xi \right)$ and $\mathcal{P}_a
\left(\bar\pi\tilde\sigma^a \pi \right)$. Therefore, it is required
to make such restrictions on the background geometry that allow to
form in the right-hand sides of \p{PB-ll-T0}, \p{PB-ll0-T0} only the
terms with above contractions. Otherwise, it will be necessary to
introduce new constraints that are linear in $\mathcal{P}_a$, and
the resulting theory will not have the correct flat limit.

It is easy to see that the needed terms  $\mathcal{P}_a
\left(\xi\sigma^a \bar\xi \right)$ and $\mathcal{P}_a
\left(\bar\pi\tilde\sigma^a \pi \right)$ will appear on the
right-hand sides of \p{PB-ll0-T0}, \p{PB-tll0-T0} in the case when
the curvature tensor \p{D-cur} of background geometry has the form
 \be \label{R-dSitter} {R}_{ab}{}^{cd} \ = \ \kappa
\left(\delta_a^c\delta_b^d-\delta_a^d\delta_b^c\right),
 \ee
where $\kappa$ is a constant. As well know, the relation
\p{R-dSitter} defines the constant curvature space-time. At zero,
positive or negative $\kappa$ we get the Minkowski, de Sitter and
anti de Sitter spaces respectively.

Under the restriction \p{R-dSitter}, the first terms in the
right-hand sides of \p{PB-1F0-l}, \p{PB-1F0-tl} are proportional
to $\left(\xi\sigma^a \bar\xi \right)\mathcal{P}_a$ and
$\left(\bar\pi\tilde\sigma^a \pi \right)\mathcal{P}_a$. To see this,
we use the identities
$$
(\sigma^a)_{\alpha\dot\alpha}(\sigma_{ab})_{\beta}{}^{\gamma}=-(\sigma_b)_{\beta\dot\alpha}\delta_{\alpha}^{\gamma}
+\frac12\,(\sigma_b)_{\alpha\dot\alpha}\delta_{\beta}^{\gamma}\,,\qquad
(\sigma^a)_{\alpha\dot\alpha}(\tilde\sigma_{ab})^{\dot\beta}{}_{\dot\gamma}=(\sigma_b)_{\alpha\dot\gamma}\delta^{\dot\beta}_{\dot\alpha}
-\frac12\,(\sigma_b)_{\alpha\dot\alpha}\delta^{\dot\beta}_{\dot\gamma}\,,
$$
and obtain
 \be \left(\xi\sigma^a \bar\xi
\right)M_{ab}=-\frac12\left(\xi\sigma_b \bar\xi \right)K\,,\qquad
\left(\bar\pi\tilde\sigma^a \pi
\right)M_{ab}=\frac12\left(\bar\pi\tilde\sigma_b \pi \right)K\,.
 \ee
It is also well known that for the curvature tensor \p{R-dSitter} of
constant curvature spaces the following relation takes place
 \be
\label{zero} \mathcal{D}_h {R}_{ab}{}^{cd} = 0.
 \ee
Therefore the second terms in the right-hand sides of \p{PB-1F0-l}
and \p{PB-1F0-tl} are nullified.

Note that due to the equalities
$$
(\sigma^{ab})_{\alpha}{}^{\beta}(\sigma_{ab})_{\gamma}{}^{\delta}=-2\delta_{\alpha}^{\delta}\delta_{\gamma}^{\beta}
+\delta_{\alpha}^{\beta}\delta_{\gamma}^{\delta} \,,\qquad
(\tilde\sigma^{ab})^{\dot\alpha}{}_{\dot\beta}(\tilde\sigma_{ab})^{\dot\gamma}{}_{\dot\delta}=
-2\delta^{\dot\alpha}_{\dot\delta}\delta^{\dot\gamma}_{\dot\beta}
+\delta^{\dot\alpha}_{\dot\beta}\delta^{\dot\gamma}_{\dot\delta}\,,
\qquad
(\sigma^{ab})_{\alpha}{}^{\beta}(\tilde\sigma_{ab})^{\dot\gamma}{}_{\dot\delta}=0\,,
$$
for $\sigma$-matrices, in the case of spaces of constant curvature
\p{R-dSitter}, the second term in \p{ell0-1mod} takes the form: \be
\label{RMM-dS} -  \frac{1}{4}\,{R}_{ab}{}^{cd}
M^{ab}M_{cd}=\frac{1}{2}\,{\kappa}\left(N^2+ \bar N^2 \right). \ee

Thus, we see that in the (anti-) de Sitter space \p{R-dSitter} the
quantities \be \ell_0-  \frac{1}{4}\,{R}_{ab}{}^{cd}
M^{ab}M_{cd}\,,\qquad \ell\,,\qquad\tilde\ell\,, \ee form the closed
algebra with respect of Poisson brackets: \bea \label{PB-ll-T00}
\left\{ \ell ,\tilde\ell \right\} & = & - K \left( \ell_0-
\frac{1}{4}\,{R}_{ab}{}^{cd} M^{ab}M_{cd}\right), \\ [6pt]
\label{PB-ll0-T00} \left\{ \ell_0-  \frac{1}{4}\,{R}_{ab}{}^{cd}
M^{ab}M_{cd}\ ,\ \ell \right\} & = & -2\kappa K\ell \,,
\\ [6pt]
\label{PB-tll0-T00} \left\{ \ell_0-  \frac{1}{4}\,{R}_{ab}{}^{cd}
M^{ab}M_{cd} \ , \ \tilde\ell \right\} & = & 2\kappa K\tilde\ell \,.
\eea

But in this closed algebra there are only quantities $\ell$ and
$\tilde\ell$. At the same time, taking into account the flat
constraints \p{const-sp-1} and \p{const-sp-2}, the curved
counterparts of these constraints should contain terms with constant
$\bm{\mu}$.
Therefore, these quantities can no be considered as curved space curved counterparts of flat constraints since they have no correct flat limit.

\subsection{Final solution to the constraints}

We move on to final constructing the curved space constraints. First of all, a natural step to consider, as the constraints, the expressions
$\ell-\bm{\mu}\approx0$,\, $\tilde\ell-\bm{\mu}\approx0$ and $\ell_0-  \frac{1}{4}\,{R}_{ab}{}^{cd} M^{ab}M_{cd}\approx0$. However, taking into account the relations \p{PB-ll0-T00}, \p{PB-tll0-T00},
one sees that the Poisson brackets among them do not form a closed algebra. We will show that the quantities $\ell-\bm{\mu}$,\, $\tilde\ell-\bm{\mu}$ and $\ell_0-  \frac{1}{4}\,{R}_{ab}{}^{cd} M^{ab}M_{cd}$ can be modified in such a way to obtain the true constraints $\mathcal{F}_0,\, \mathcal{F}, \, \mathcal{\tilde{F}}$
forming the closed algebra and having correct flat limit.

Let us begin with modification of the quantity $\ell_0-  \frac{1}{4}\,{R}_{ab}{}^{cd} M^{ab}M_{cd}$ requiring that the modified quantity commute with
$\ell-\bm{\mu}$ and $\tilde\ell-\bm{\mu}$. Using Poisson brackets
\be
\left\{ N,\ell \right\}=\left\{ \bar N,\ell \right\}=\frac12 \left\{ K,\ell \right\}=-\ell\,,\qquad
\left\{ N,\tilde\ell \right\}=\left\{ \bar N,\tilde\ell \right\}=\frac12 \left\{ K,\tilde\ell \right\}=\tilde\ell
\ee
and \p{PB-ll0-T00}, \p{PB-tll0-T00}, we see that the quantity, which is equal
to the sum of $\displaystyle \ell_0-  \frac{1}{4}\,{R}_{ab}{}^{cd} M^{ab}M_{cd}$ plus the additional term
\be\lb{add-terms}
-\frac{{\kappa}}{2}\,K^2 \qquad\quad \mbox{or}\qquad\quad -2{\kappa}N\bar N \qquad\quad \mbox{or}\qquad\quad -{\kappa}(N^2+\bar N^2)\,,
\ee
has zero Poisson brackets with $\ell-\bm{\mu}$ and $\tilde\ell-\bm{\mu}$.
However, at the constraint $\mathcal{U}=\ N -\bar N \approx0$ ,
the following equalities take place
\be
-\frac{{\kappa}}{2}\,K^2 = -2{\kappa}N\bar N = -{\kappa}(N^2+\bar N^2)\,
\ee
that allows us to shorten a number of the variants in \p{add-terms}. We choose the first option and define the quantity
\be\lb{L0-a}
{\mathcal{F}}_{0, \, \gamma} \ := \ \ell_0-  \frac{1}{4}\,{R}_{ab}{}^{cd} M^{ab}M_{cd}-\frac{{\kappa}}{2}\,K^2 +\gamma\  \approx \ 0 \,
\ee
as the true modified constraint with arbitrary constant parameter $\gamma$ which will be fixed later.
Let us emphasize once again the important property obtained, the constraint \p{L0-a} has
zero Poisson brackets with $\ell-\bm{\mu}$ and $\tilde\ell-\bm{\mu}$:
\be\lb{F0-ll}
\left\{ {\mathcal{F}}_{0, \, \gamma},\ell \right\}=\left\{ {\mathcal{F}}_{0, \, \gamma},\tilde\ell \right\}=0 \,.
\ee
We will show that this quantity can be associated with true constraint $\mathcal{F}_{0}$ at some fixed $\gamma.$

Let us turn to the constraints $\ell-\bm{\mu}\approx0$, $\tilde\ell-\bm{\mu}\approx0$. The relation \p{PB-ll-T00} shows that their Poisson bracket can not reproduce the
quantity $\mathcal{F}_{0, \, \gamma}$ at any $\gamma$. Therefore, if we desire to obtain the true constraints $\mathcal{F}$ and $\mathcal{\tilde{F}}$ from
the quantities $\ell-\bm{\mu}$ and $\tilde\ell-\bm{\mu}$, last should be modified in such a way
that their Poisson bracket is equal to a linear combination of the new constraints, unlike the Poisson bracket \p{PB-ll-T00}.

The desired modification is realized by adding
the terms containing $K^2$ to the constraints  $\ell-\bm{\mu}\approx0$, $\tilde\ell-\bm{\mu}\approx0$. As a result, we arrive to the following functions
\be\lb{LL-a}
{\mathcal{F}_{\beta_1}} \ := \ \ell-\bm{\mu} +\beta_1 K^2 \  \approx \ 0 \,,\qquad\quad
\tilde{\mathcal{F}_{\beta_2}} \ := \ \tilde\ell-\bm{\mu} +\beta_2 K^2 \  \approx \ 0 \,,
\ee
where $\beta_1$ and  $\beta_2$ are some arbitrary constants.
In the general case, these constants are not equal to each other, $\beta_1\neq\beta_2$, since the constraints \p{LL-a} are real and are not related one to another.

Now we require that the quantities $\mathcal{F}_{0, \, \gamma}, \, \mathcal{F}_{\beta_1}, \, \mathcal{F}_{\beta_2}$ form the closed algebra. We will show that this requirement completely fix the constants $\gamma, \, \beta_1, \, \beta_2.$

Since $\left\{M_{ab},K \right\}=0$, the constraint \p{L0-a}
have zero Poisson brackets with constraints \p{LL-a}:
$\left\{ {\mathcal{F}}_{0,\, \gamma},\mathcal{F}_{\beta_1} \right\}=\left\{ {\mathcal{F}}_{0, \, \gamma},\tilde{\mathcal{F}_{
\beta_2}}  \right\}=0$. The Poisson bracket of the quantities \p{LL-a} equals
\be\lb{alg-LL-a}
\left\{{\mathcal{F}_{\beta_1}},\tilde{\mathcal{F}_{\beta_2}}  \right\}=-K{\mathcal{F}}_0+4 K\left(\beta_2 {\mathcal{F}_{\beta_1}}+\beta_1\tilde{\mathcal{F}_{\beta_2}}  \right)
-\frac12\left({\kappa}+16\beta_1\beta_2  \right)K^3+
\left(\gamma+4\bm{\mu}\left(\beta_1+\beta_2\right)\right)K\,.
\ee
For closure of this algebra, the third and fourth terms on the right-hand side must be equal to zero.
This leads to the following conditions on the constants $\gamma$, $\beta_1$ and  $\beta_2$:
\be\lb{val-b}
\beta_1\beta_2=-\frac1{16} \, {\kappa}\,,\qquad \gamma=-4\bm{\mu}\left( \beta_1+\beta_2\right)\,.
\ee
Further we consider two cases, ${\kappa}<0$ (anti de Sitter space) and ${\kappa}>0$ (de Sitter case).

\textbf{Case of anti de Sitter space.} In this case the simplest solution of the equations \p{val-b} is
\be
\label{AdS}
\beta_1=\beta_2=\beta=\frac14 \, |{\kappa}|^{1/2}\,,\qquad \gamma=-2\mu|{\kappa}|^{1/2}\,.
\ee
Denoting $\mathcal{F}_0 = \mathcal{F}_{0, \, \gamma=-2\bm{\mu}|{\kappa}|^{1/2}}$, \, $\mathcal{F}=\mathcal{F}_{\beta_{1}= \frac14 \, |{\kappa}|^{1/2}}$, \,
$\mathcal{\tilde{F}}=\mathcal{\tilde{F}}_{\beta_{2}= \frac14 \, |{\kappa}|^{1/2}}$, we obtain the final constraints for the case of AdS space
\begin{eqnarray}
{\mathcal{F}}_0 &=&
\mathcal{P}_a \mathcal{P}^a-  \frac{1}{4}\,{R}_{ab}{}^{cd} M^{ab}M_{cd}
-\frac{1}{2}\,{\kappa}K^2 -2\bm{\mu}|{\kappa}|^{1/2} \ \approx \ 0 \,, \label{const-sp-b}\\ [5pt]
{\mathcal{F}} &=&
\mathcal{P}_a \left(\xi\sigma^a \bar\xi \right)-\bm{\mu}+\frac14 \, |{\kappa}|^{1/2} K^2 \ \approx \ 0  \,,  \label{const-sp-1-b}\\ [5pt]
\tilde{\mathcal{F}} &=&
\mathcal{P}_a \left(\bar\pi\tilde\sigma^a \pi \right) -\bm{\mu}+\frac14 \, |{\kappa}|^{1/2} K^2 \ \approx \ 0  \,,  \label{const-sp-2-b}\\ [5pt]
\mathcal{U} &=& N -\bar N \ \approx \ 0 \,.  \label{const-sp-3-b}
\end{eqnarray}
The constraints \p{const-sp-b} and \p{const-sp-3-b} have zero Poisson brackets with other constraints:
\be\lb{alg-L0U-L}
\left\{ {\mathcal{F}}_0,{\mathcal{F}}  \right\}=\left\{ {\mathcal{F}}_0,\tilde{\mathcal{F}}  \right\}=
\left\{ \mathcal{U},{\mathcal{F}}  \right\}=\left\{ \mathcal{U},\tilde{\mathcal{F}}  \right\}=
\left\{ {\mathcal{F}}_0,\mathcal{U}  \right\}=0\,.
\ee
The only non-zero Poisson bracket of constraints \p{const-sp-b}-\p{const-sp-3-b} has the form
\be\lb{alg-LtL}
\left\{{\mathcal{F}},\tilde{\mathcal{F}}  \right\}=-K{\mathcal{F}}_0+|{\kappa}|^{1/2} K\left({\mathcal{F}}+\tilde{\mathcal{F}}  \right)\,.
\ee
We see that the quantities \p{const-sp-b}-\p{const-sp-3-b} form a system of the first class constraints as it is expected.
Note that taking into account \p{K} and \p{RMM-dS}, the constraint \p{const-sp-b} is represented in the form:
\be
{\mathcal{F}}_0 = \mathcal{P}_a \mathcal{P}^a-  {\kappa}N\bar N -2\mu|{\kappa}|^{1/2} \ \approx \ 0 \,.
\ee

\textbf{Case of de Sitter space.}  In this case, the simplest solution of the equations \p{val-b} is
\be
\beta_1=-\beta_2=\beta=\frac14 \, {\kappa}^{1/2}\,,\qquad \gamma=0\,.
\ee
That is, in the case of the background de Sitter space, it is possible to select constraints with zero  constant $\gamma$.
Denoting $\mathcal{F}_0 = \mathcal{F}_{0, \, \gamma=0}$, \, $\mathcal{F}=\mathcal{F}_{\beta_{1}= \frac14 \, {\kappa}^{1/2}}$, \,
$\mathcal{\tilde{F}}=\mathcal{\tilde{F}}_{\beta_{2}= -\frac14 \, {\kappa}^{1/2}}$, we obtain the final constraints for the case of dS space
\begin{eqnarray}
{\mathcal{F}}_0 &=& \mathcal{P}_a \mathcal{P}^a-  \frac{1}{4}\,{R}_{ab}{}^{cd} M^{ab}M_{cd}
-\frac{1}{2}\,{\kappa}K^2  \ \approx \ 0 \,, \label{const-sp-d}\\ [5pt]
{\mathcal{F}} &=& \mathcal{P}_a \left(\xi\sigma^a \bar\xi \right)-\bm{\mu}+\frac14 \, {\kappa}^{1/2} K^2 \ \approx \ 0  \,,  \label{const-sp-1-d}\\ [5pt]
\tilde{\mathcal{F}} &=& \mathcal{P}_a \left(\bar\pi\tilde\sigma^a \pi \right) -\bm{\mu}-\frac14 \, {\kappa}^{1/2} K^2 \ \approx \ 0  \,,  \label{const-sp-2-d}\\ [5pt]
\mathcal{U} &=& N -\bar N \ \approx \ 0 \,.  \label{const-sp-3-d}
\end{eqnarray}
The only non-zero Poisson bracket of constraints \p{const-sp-d}-\p{const-sp-3-d} is
\be\lb{alg-LtLd}
\left\{{\mathcal{F}},\tilde {\mathcal{F}}  \right\}=-K{\mathcal{F}}_0-{\kappa}^{1/2} K\left({\mathcal{F}}-\tilde{\mathcal{F}} \right)\,.
\ee
The remaining Poisson brackets of the constraints \p{const-sp-d}-\p{const-sp-3-d} are equal to zero.
As well as in the case of the anti-de Sitter space \p{alg-L0U-L}, and the constraints \p{const-sp-d}-\p{const-sp-3-d} form a system of the first class constraints.

\vspace{0.5cm}

\setcounter{equation}{0}

\section{Summary}

Let us briefly summarize the results. In this paper we have proposed the new classical particle model in background gravitational field
with Lagrangian \p{L-sp-nf} and constraints \p{const-sp-b}-\p{const-sp-3-b} or \p{const-sp-d}-\p{const-sp-3-d}, that is a curved space time generalization of the infinite spin particle model in flat space \cite{BFIR}. The model under consideration is described by curved space-time coordinates $x^{\mu}$ and, as well as the corresponding flat space infinite spin particle model, by the additional Weyl spinor coordinates $\xi^{\alpha}, \, \bar{\xi}^{\dot{\alpha}}$ belonging to target space of the space-time. We work in the first order formalism and the main problem is to construct a system of the constraints forming the closed Poisson algebra. We begin with general enough external gravitational field with torsion and show that conditions of closure of the constraint algebra uniquely fix the background to be a constant curvature space. The constraints are derived in the explicit form both for AdS and for dS spaces. In the flat space limit we return back to the model proposed in \cite{BFIR}.The result is consistent with modern understanding that the true vacuum for higher spin fields is AdS space (see, e.g., the reviews \cite{Vas2001,Vas2002,Vas2005,BBS}).

\end{document}